\documentclass[9pt,twocolumn,twoside,lineno]{pnas-new}
\usepackage{multirow}
\usepackage{makecell}
\usepackage{titlesec}
\usepackage{xcolor}
\setcellgapes{4pt}

\templatetype{pnasresearcharticle} 

\title{Communication by means of Modulated Johnson Noise}

\author[a]{Zerina Kapetanovic}
\author[b]{Miguel Morales}
\author[a,c]{Joshua R. Smith} 

\affil[a]{Department of Electrical and Computer Engineering, University of Washington, Seattle, WA}
\affil[b]{Department of Physics, University of Washington, Seattle, WA}
\affil[c]{Department of Computer Science and Engineering, University of Washington, Seattle, WA}

\leadauthor{Kapetanovic} 

\authorcontributions{Z.K. and J.S. contributed conceptualization and methodology. Z.K., M.M., and J.S. contributed to analyzing the physics of the system. Z.K contributed to investigation, visualization, and original draft writing. J.S contributed to funding acquisition, supervision, and final draft review and editing. }
\correspondingauthor{\textsuperscript{2}To whom correspondence should be addressed. E-mail: zerinak@uw.edu}


\begin{abstract}
We present the design of a new passive wireless communication method that does not
rely on ambient or generated RF sources. Instead, we exploit the Johnson (thermal) noise of a resistor to transmit information bits wirelessly. By switching the load connected to an antenna between a resistor and an open or short circuit, we can achieve data rates of up to 26bps and distances of up to 7.3 meters. This communication method is orders of magnitude less power consuming than conventional communication schemes and presents the opportunity to enable wireless communication for power and connectivity-constrained applications.
\end{abstract}

\begin{document}

\maketitle
\thispagestyle{firststyle}
\ifthenelse{\boolean{shortarticle}}{\ifthenelse{\boolean{singlecolumn}}{\abscontentformatted}{\abscontent}}{}


\dropcap{I}n passive wireless communication, an energy-constrained data transmitter sends information by modulating an RF signal generated by an RF source that is not power constrained. Because the data transmitter does not have to generate an RF signal, the power necessary to send data is orders of magnitude less than in conventional RF communication. In Modulated Backscatter Communication, a continuous wave RF carrier is generated by a dedicated device on the high-power side of the link, and the low power side encodes data by selectively reflecting this RF signal \cite{stockman}. Ambient Backscatter is another form of passive communication that makes use of pre-existing, ambient RF signals such as those generated by broadcast TV or radio towers \cite{abc}. While the low-power of the data transmission side is attractive, both methods rely on a pre-existing RF signal. 

This paper introduces a new form of passive wireless communication, Modulated Johnson Noise, in which the signal to be modulated is the Johnson noise in an unbiased (un-powered) resistor. This scheme retains the benefits of prior passive wireless communication schemes while eliminating the need for an external RF signal. This has the potential to reduce the overall energy consumption of the system, to allow more stealthy and low-interference operation, and to allow operation in areas where no ambient RF signals are available.

\begin{figure}[!ht]
    \centering
    \includegraphics[width=0.75\linewidth]{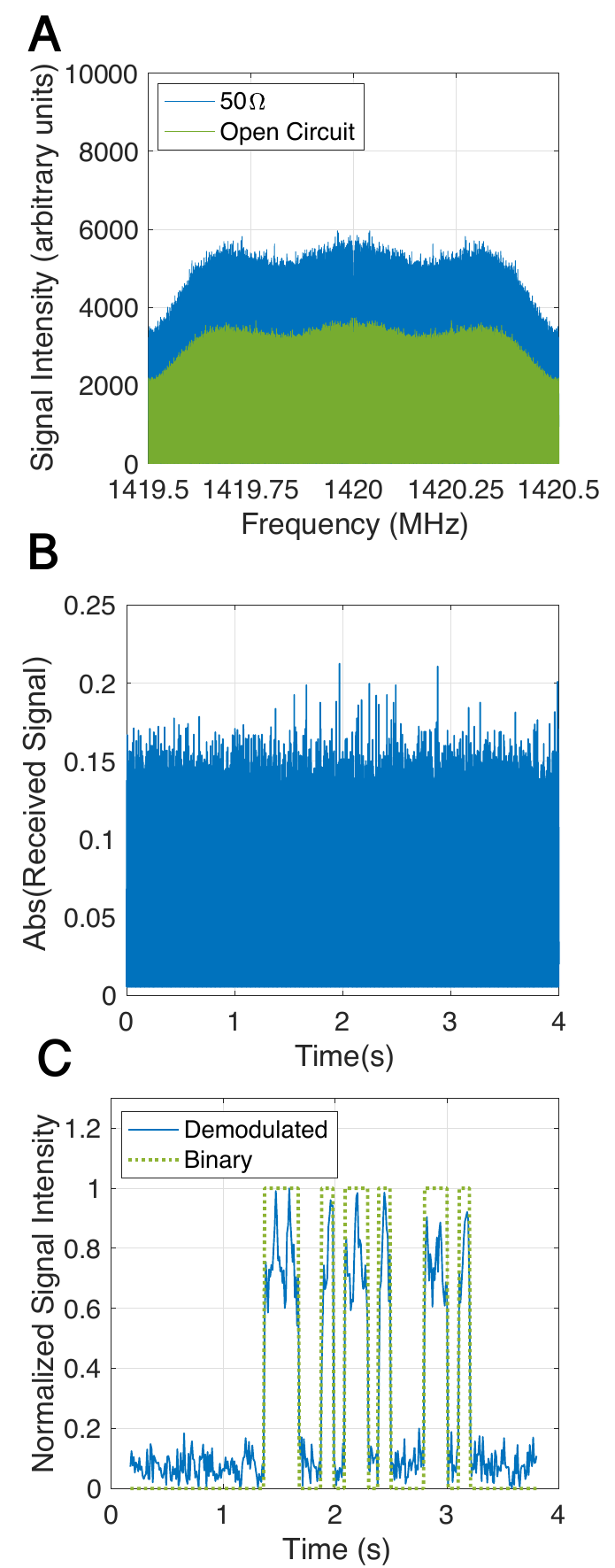}
    \caption{\textbf{Johnson Noise Communication.} Wireless communication can be enabled by modulated Johnson noise. \textbf{(A)} shows the measured frequency spectrum at 1.42GHz, which compares measurements when a 50$\Omega$ load is connected to a receiver and an open circuit load. \textbf{(B)} shows what a received signal looks like when wirelessly transmitting a data packet by modulated Johnson noise and \textbf{(C)} shows the received data packet after demodulation.}
    \label{fig:pwrmeasurements}
\end{figure}

Consider the frequency spectrum measurements shown in Fig.~\ref{fig:pwrmeasurements}\textcolor{blue}{A}, which shows the measured signal of a 50$\Omega$ terminator and open circuit terminator connected to the input of a receiver. We can see that there is a clear difference between the two measurements, which can be exploited to enable wireless communication. By selectively connecting and disconnecting an impedance matched 50$\Omega$ resistor to an antenna, information bits can be wirelessly transmitted. For example, in Fig.~\ref{fig:pwrmeasurements}\textcolor{blue}{B} we show the received signal of a data packet that was wirelessly transmitted by modulating Johnson noise. While this looks like a noisy signal, after performing demodulation, the data packet can be extracted as shown in Fig.~\ref{fig:pwrmeasurements}\textcolor{blue}{C}. While thermal noise communication has been proposed from a theoretical perspective \cite{kish2005stealth}, we present the first system that enables wireless communication by modulating Johnson noise. In this paper, we discuss the design and experimental implementation of the system and evaluate the overall performance of the wireless communication scheme that relies on modulated Johnson noise. The contributions of the paper are summarized below:

\begin{itemize}
    \item We introduce wireless communication by means of modulated Johnson noise, the first wireless system that enables devices to communicate without reliance on generated or ambient RF signals. 
    \item We present the designs and prototype of hardware that enables Johnson noise communication. Moreover, we demonstrate the transmitter can be designed to be battery-free.
    \item We evaluate the performance of the wireless system in terms of achievable throughput and communication range. The performance evaluation shows that data rates of up to 26bps can be achieved at distances of up to 7.3 meters. 
\end{itemize}

\section*{Overview of Johnson Noise}
Johnson noise is caused by the thermal vibrations of charge carriers inside of an electrical conductor (e.g., resistor) and is characterized by its mean-squared voltage, 

\begin{equation}
    V_n^2  = 4kTBRe(Z)
    \label{eq:vn}
\end{equation}

where $k$ is Boltzmann's constant, T is temperature, B is bandwidth, and Re(Z) is the the real part of the electrical conductor's impedance~\cite{thermometry}. A resistor with Johnson noise can be modeled as a Thevenin equivalent circuit that includes a noiseless resistor and a noise voltage generator with voltage given by Eq. ~\ref{eq:vn}. With a matched load resistor connected, the maximum noise power provided by the noisy resistor is 

\begin{equation}
    P_n = \frac{V_n^2}{4R} = kTB
\end{equation}
which is independent of resistance~\cite{pozar}. The thermal noise power is a function of temperature and bandwidth. As an example, a resistor at room temperature (296K) with a system bandwidth of 500MHz would result in $P_n$ = -86.9 dBm. We also note that Johnson noise is a white noise noise source and therefore independent of frequency (within the system's bandwidth). Since it is white, Johnson noise has a Gaussian distribution.

\begin{figure*}[t!]
    \centering
    \includegraphics[width=1\textwidth]{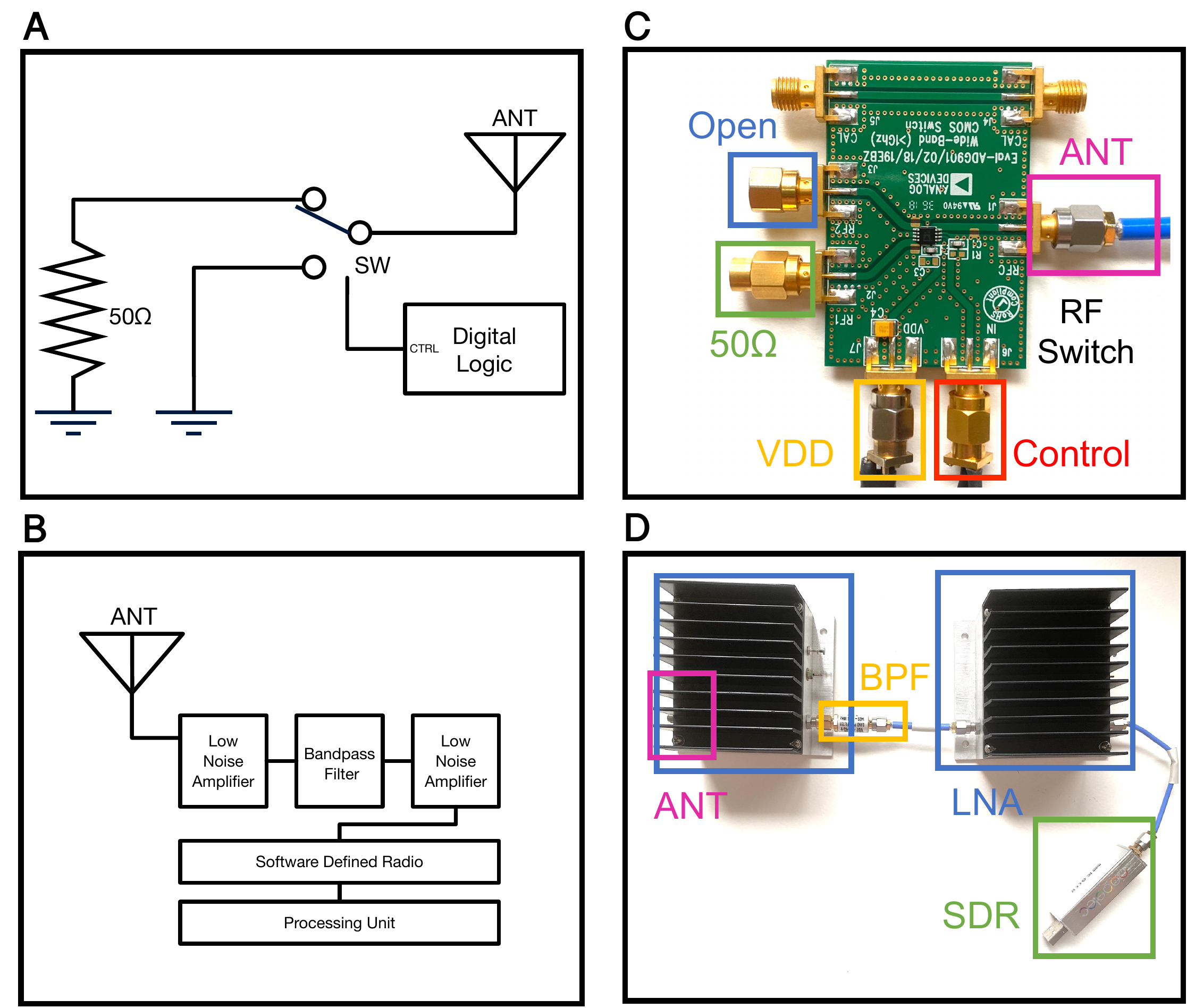}
    \caption{\textbf{Design and Implementation.} \textbf{(A)} shows the transmitter design, which switches between a 50$\Omega$ load and short circuit (or open circuit) to modulated information bits. \textbf{(B)} shows the receiver, which is composed of two low noise amplifiers (LNA) who's output is fed into a software defined radio. The bandpass filter between the two LNAs is added to prevent feedback oscillation and the data received by the SDR can be processed by laptop PC or small-form factor computer (e.g. Raspberry Pi). \textbf{(C)} and \textbf{(D)} show the prototype implementation of the transmitter and receiver, respectively.} 
    \label{fig:hardware}
\end{figure*}

\section*{Design and Implementation} 
We design and implement a transmitter (TX) and receiver (RX) to enable wireless communication by modulating Johnson noise. First, the transmitter requires an RF switch and a processing unit to control switching between an open or short circuit, and resistor load connection shown in Fig.~\ref{fig:hardware}\textcolor{blue}{A}. On the receive side, shown in Fig.\ref{fig:hardware}\textcolor{blue}{B}, two high gain, low noise amplifiers (LNA) amplify the very low-power signals being transmitted while maintaining a good signal-to-noise ratio. Between the two LNAs is a bandpass filter to prevent feedback oscillation. The entire system is designed to operate at 1.42GHz with 50$\Omega$ impedance. In Fig.~\ref{fig:hardware}\textcolor{blue}{C} and ~\ref{fig:hardware}\textcolor{blue}{D}  we show the prototype implementation of the transmitter and receiver. The transmitter uses an RF switch and switches between a 50$\Omega$ RF terminator (with 50$\Omega$ impedance) and an open circuit terminator, both of which are very well shielded~\cite{adg901}. On the receive side there are two LNAs with combined gain of approximately 84dB, a 50MHz bandpass filter, and a software defined radio (SDR)~\cite{lna, SDR, bpf}. We constructed pyramidal horn antennas for both the TX and RX side. The antennas were characterized to have approximately 13.6dBi of gain. 

\subsection*{Data Encoding and Modulation}
Since the transmitter is switching between two states, data can be modulated by performing ON-OFF keying. A 0-bit is transmitted by simply staying in the OFF state (open circuit) and a 1-bit is transmitted by switching between both ON and OFF states (open circuit and 50$\Omega$) using a square-wave subcarrier frequency. Mathematically, the transmitted signal is written as,

\begin{equation}
    s_{rx} = m\cdot sgn\big(sin(2\pi f_{sc}t)\big)
\end{equation}

where m represents the bit to be encoded and takes on a 0 or 1 value. The subcarrier frequency is defined as $f_{sc}$. 

\subsection*{Packet Detection and Demodulation}
Our system assumes that the $f_{sc}$ is known on the RX side, which allows us to perform heterodyne detection. Our demodulation is inspired by techniques used in radio astronomy, in particular, the Dicke radiometer~\cite{dicke}. We can view our entire system as a distributed Dicke radiometer, where the switching between two states occurs on the TX side, while the amplification and integration occurs on the RX side. The receiver performs heterodyne detection which allows the system to reject noise by filtering out any received signal power that is outside the narrow bandwidth of the subcarrier. In other words, the receiver generates the same square-wave subcarrier signal with amplitude values of +1 and -1, multiplies the received signal with this receive-side subcarrier, and accumulates (integrates) the product values for a duration that is less than or equal to that of the duration of one bit. This process results in a demodulated signal intensity and is given by,

\begin{equation}
    \langle s_{rx},s_{sc} \rangle = \int_{0}^{T} s_{rx}(t)s_{sc}(t) dt 
\end{equation}

where $T$ is the total integration time. The demodulated signal intensity is compared to a threshold value to determine whether a 1 or 0-bit was transmitted. In Fig.~\ref{fig:pwrmeasurements}\textcolor{blue}{B} we show an example of a received data packet before demodulation that was wirelessly transmitted, and in Fig.~\ref{fig:pwrmeasurements}\textcolor{blue}{C} we show the data packet after demodulation and its binary format after performing thresholding.

\section*{System Validation}
With this method of wireless communication comes two key questions that must first be answered: (1) Are we indeed modulating Johnson noise and not another source (e.g., ambient RF or control signal feedthrough)?  (2) How does modulated Johnson noise work if the entire system is at thermal equilibrium? 

\begin{figure*}[t!]
    \centering
    \includegraphics[width=1\textwidth]{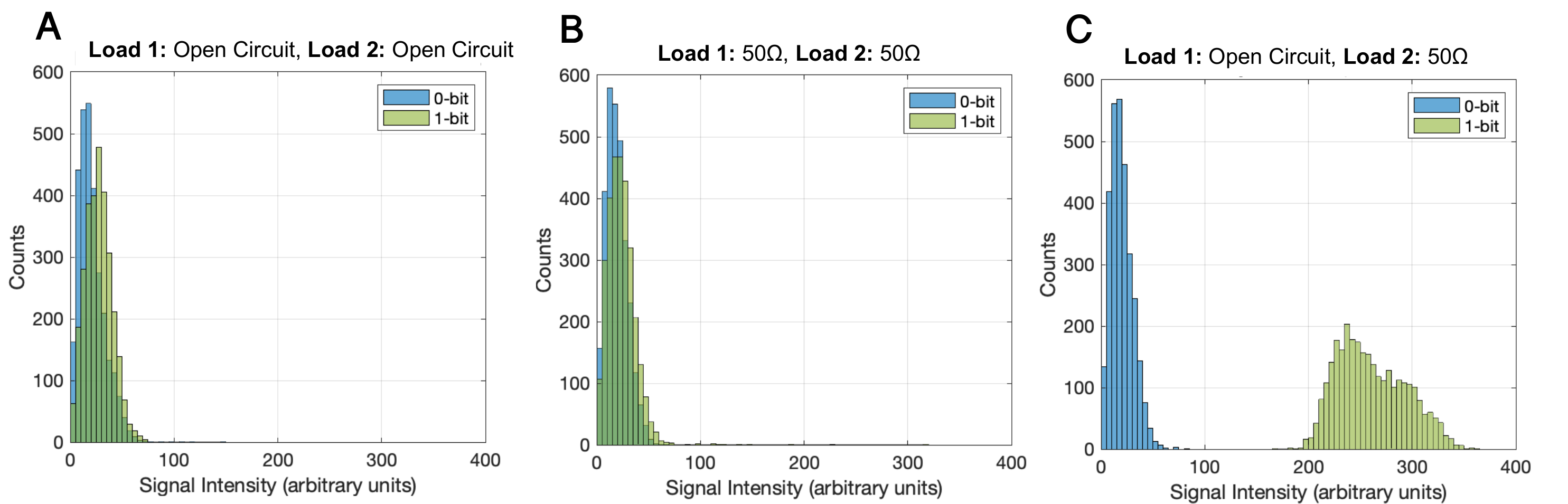}
    \caption{\textbf{Feedthrough Evaluation.} To ensure the system has proper signal isolation, we evaluate the possibility of feedthrough from the control signal used for the RF switch in the transmitter. In \textbf{(A)} the histogram distribution of 0 and 1-bit transmissions after demodulation when switching between two identical open circuit loads is shown. Similarly, \textbf{(B)} shows the results when switching between to identical 50$\Omega$ loads at room temperature. In \textbf{(A)} and \textbf{(B)}, the 0 and 1-bits are clearly not distinguishable. However, when switching between an open circuit and 50$\Omega$ load, shown in \textbf{(C)}, there is a clear difference between 0 and 1-bit transmissions.}
    \label{fig:feedthrough}
\end{figure*}

\subsection*{Evaluating Feedthrough}
One concern that comes up with this system is whether there is feedthrough from the control signal used for the RF switch. We have performed control experiments to evaluate signal isolation. If the system is truly using thermal noise, then if it is configured to switch between two open circuit loads or two 50$\Omega$ loads at room temperature, it should not be possible to decode data transmissions. We evaluate these two scenarios and compare the results to that of switching between 50$\Omega$ and an open circuit.  

An additional potential concern is that we might not truly be using Johnson noise. Perhaps some pre-existing RF signal, such as 60Hz, broadcast TV, or even galactic emissions from neutral hydrogen, are being picked up in our apparatus, and then being modulated by our two different load resistances, resulting in different outputs from our TX antenna. To control for this, we tested our data transmission method by switching between two 50$\Omega$ loads at drastically different temperatures inside an anechoic chamber. If we truly are using Johnson noise, and not some other mechanism related to the different impedances, then two identical impedances at different temperatures should also work. We demonstrate that data can be modulated by switching between a 50$\Omega$ load at room temperature (296K) and a 50$\Omega$ load submerged in liquid nitrogen (77K). 

\subsubsection*{(1) Control Experiment: Feedthrough} A cabled benchtop experiment is set up to evaluate the three different scenarios.  Data bits (1 and 0) are transmitted and then demodulated by the receiver. The demodulated data is presented in the form of histogram distributions which show the signal intensity of demodulated 1 and 0-bits. The open circuit and 50$\Omega$ loads used were RF shielded terminators. In Fig.~\ref{fig:feedthrough}\textcolor{blue}{A} we show the system evaluated when switching between two open circuit loads. In Fig.~\ref{fig:feedthrough}\textcolor{blue}{B} we show the system evaluated when switching between two 50$\Omega$ loads. We compare this to the results shown in Fig.~\ref{fig:feedthrough}\textcolor{blue}{C}, where the system is switching between open circuit and 50$\Omega$. From these results, we can see that when switching between two open circuit states or two 50$\Omega$ states, the distribution of 1 and 0-bits look nearly identical. On the other hand, when switching between open circuit and 50$\Omega$ this is a clear difference between 1 and 0-bit transmissions.

\subsubsection*{(2) Control Experiment: Temperature modulation} 
\begin{figure}[h!]
    \centering
    \includegraphics[width=0.8\columnwidth]{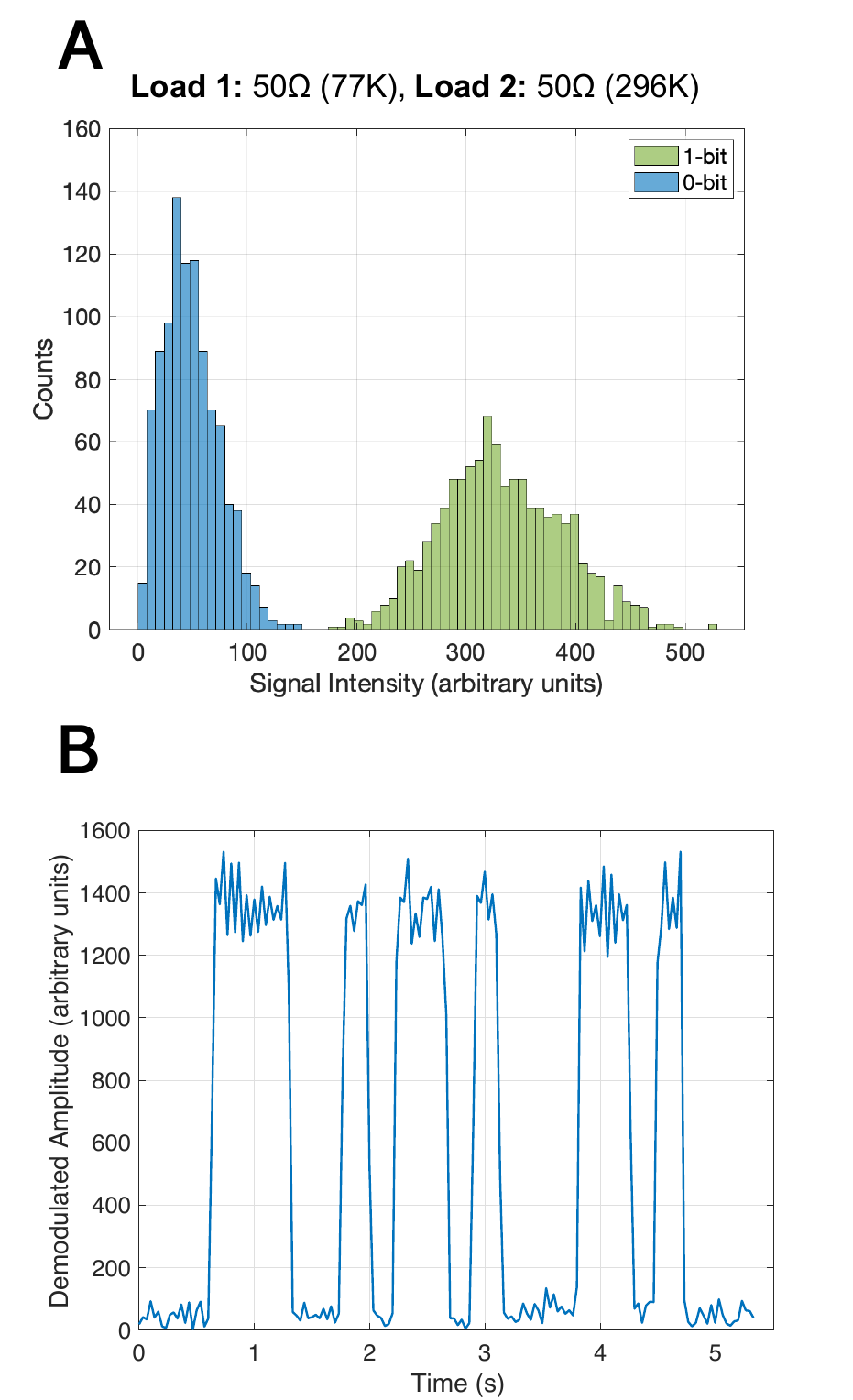}
    \caption{\textbf{Temperature Modulation.} Since Johnson noise power is a function of temperature, it should be possible to transmit information by switching between two 50$\Omega$ loads that are at drastically different temperatures. We modulate information bits by switching between two 50$\Omega$ loads, where one load has a temperature of 296K and the other has a temperature of 77K. \textbf{(A)} shows the histogram distribution of 1 and 0-bits after demodulation and \textbf{(B)} shows a demodulated data packet.}
    \label{fig:hotcold}
\end{figure}

Using the same hardware setup as for the feedthrough experiments, we modulate information bits by switching between two 50$\Omega$ loads that are at two different temperatures. A 50$\Omega$ load is submerged in liquid nitrogen (77K), while the other 50$\Omega$ load is at room temperature (296K). With this setup, we transmit data packets. In Fig.~\ref{fig:hotcold}\textcolor{blue}{A} we show the demodulated data and in Fig.~\ref{fig:hotcold}\textcolor{blue}{B} we show the distribution of 1 and 0-bits after demodulation. Here, data is transmitted using a data rate of 5bps and a subcarrier frequency of 100Hz. 

\subsection*{Evaluating Noise Temperature}
One may worry that if the entire experiment is done at a single temperature (e.g. room temperature) then it should not be possible to observe modulated thermal signals due to a lack of contrast between the resistor and its surroundings. The question can be refined further and there are two separate potential issues. First, the transmit side and the receive side are at the same temperature and second, the two loads on the transmit side (50$\Omega$ and open circuit) are at the same temperature as one another. In the following sections we address these concerns by first discussing the temperature contrast between TX and RX, as well as 50$\Omega$ and open circuit loads. We then present experimental measurements that estimate the noise temperature of each load and compare the measured noise of each load to a theoretical analysis. 

\subsubsection*{(1) TX-RX Contrast} If the receive side and the transmit side of the system are at the same temperature, does this mean that our method should not work? If this was the case, then by the same argument, it would not be possible to observe Johnson noise with a room temperature oscilloscope, which in fact is easily observed. Furthermore, it is straightforward to perform a benchtop experiment that is analogous to our communication scheme: if one connects or disconnects a room temperature resistor (e.g 1M$\Omega$) to an oscilloscope with a 1M$\Omega$ input impedance, one sees a much larger signal in the connected case than in the disconnected case. This is analogous to our communication set up, but in the communication scheme the scope lead has been replaced by a pair of antennas. In the disconnected case, the oscilloscope’s LNA is poorly matched, and one is left seeing just the amplifier’s internal noise. In the connected case, the noise from the resistor is much larger than the amplifier's internal noise.

Active amplifiers, such as those found in an oscilloscope or the low noise amplifiers (LNA) in our receiver, are not in thermal equilibrium and are characterized by a noise temperature ($T_N$) that is "cold": lower than the physical temperature ($T_P$) of the apparatus itself.  This means that the radiation emitted by the receive antenna coupled to the LNA is equivalent to that of a colder 50$\Omega$ resistor.


\begin{figure*}[t!]
\begin{minipage}{.48\textwidth}
  \centering
  \includegraphics[width=0.8\columnwidth]{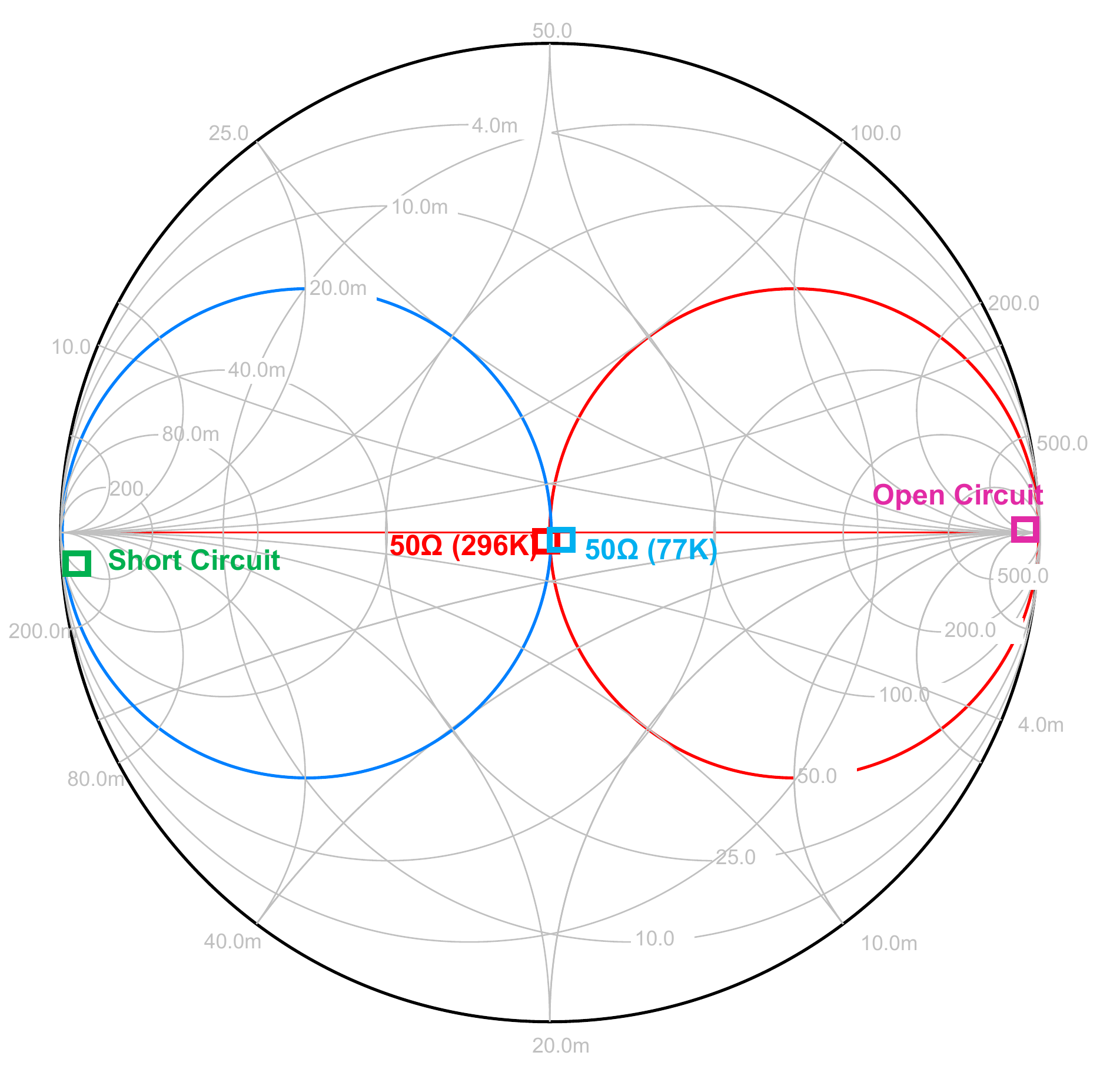} 
  \caption{\textbf{Load Impedance.} A Smith chart showing the impedance of each evaluated load. The open and short circuit are both mismatched and have a resistance of approximately 17k$\Omega$ and 0.25$\Omega$, respectively. The 50$\Omega$ load is very well matched and has a 50$\Omega$ impedance at both 296K and 77K.}
  \label{fig:smithchart}
\end{minipage}%
\hfill
\begin{minipage}{.48\textwidth}
  \centering
  \includegraphics[width=1\columnwidth]{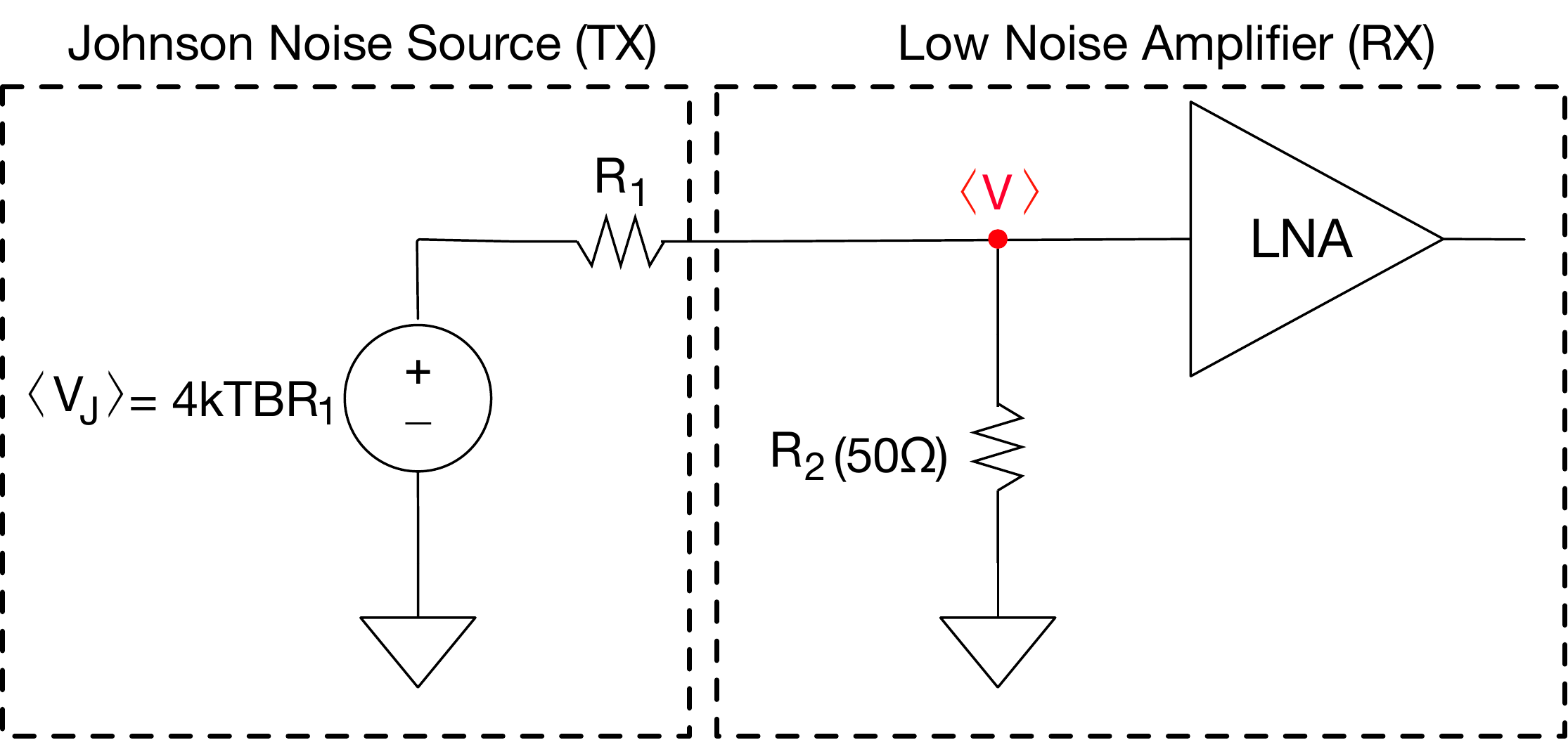}
  \vspace{48pt}
  \caption{\textbf{Equivalent Circuit.} The equivalent circuit model for a cabled scenario showing the voltage divider formed by a noisy resistor and LNA input impedance.}
  \label{fig:circuit}
\end{minipage}
\end{figure*}

\subsubsection*{(2) 50$\Omega$-Open Contrast} The two loads we switch between for signaling are at the same temperature. Furthermore, the Johnson noise {\em power} is independent of the value of the resistance. Thus one might worry that no useful contrast can be generated. The Johnson noise {\em mean squared voltage} on the other hand is proportional to the value of the resistance. We analyze this quantitatively below, but in summary this is because the 50$\Omega$ load is well matched to the LNA input, and the open and short are poorly matched. 


To evaluate this explanation quantitatively, we first measure the impedance of the open, short, and 50$\Omega$ loads at our operating frequency of $1.42$GHz. Fig.~\ref{fig:smithchart} shows the impedance of the open, short, and 50$\Omega$ loads. As expected the open circuit has a high resistance, approximately 17k$\Omega$, and the short circuit has a low resistance, approximately 0.25$\Omega$. To predict the mean squared Johnson noise voltage across the load resistor, we plug these measured impedance values into the expression for the mean squared Johnson noise voltage defined by Eq.~\ref{eq:vn}. Now, we need to model the effect of the LNA, in particular the effect of its input impedance on the measured signal. The input impedance of the LNA is 50$\Omega$, meaning that it can be modeled as an equivalent circuit with a 50$\Omega$ resistor to ground and an infinite input impedance amplifier observing the voltage at the junction between the LNA's input and this 50$\Omega$ shunt resistor. Fig.~\ref{fig:circuit} shows the voltage divider formed by the noise resistor and LNA input. The mean squared voltage observed by the amplifier, $\langle V^2 \rangle$, is given by

\begin{equation}
    \langle V^2 \rangle = \langle V_J^2 \rangle \cdot g^2 = 4kTBR_1 \cdot g^2
    \label{eq:v2_th}
\end{equation}
where 
\begin{equation}
    g = \frac{R_2}{R_1 + R_2}
    \label{eq:vgain}
\end{equation}


is the gain of the voltage divider. As shown in Fig.~\ref{fig:smithchart}, $R_1$ represents the resistance of the Johnson noise source and $R_2$ = 50$\Omega$ is the LNA's input impedance. The infinite input impedance amplifier block in Fig.~\ref{fig:circuit} measures the voltage $V$ at the node where $R_1$ and $R_2$ connect. The mean squared input voltage to the amplifier is  given by Eq.~\ref{eq:v2_th}. Expressed as multiples of $kTB$, the mean squared noise voltages $ \langle V^2 \rangle$ for our open, short, and 50$\Omega$ loads are predicted to be

\begin{equation*}
      \langle V_{open}^2 \rangle  = 0.58 \cdot kTB
\end{equation*}

\begin{equation*}
      \langle V_{short}^2 \rangle = 1.0 \cdot kTB
\end{equation*}

\begin{equation*}
      \langle V_{50}^2 \rangle = 50 \cdot kTB
\end{equation*}

The varying constants in front of $kTB$ demonstrate that the impedance matching differences lead to small signals from the open and short, and larger signals from the 50$\Omega$ load. Later (after addressing calibration) we provide a detailed comparison of predicted mean squared noise voltages to experimentally measured values for each of the loads of interest, shown in Table~\ref{table:v2calc}. Note that the experimentally observed contrast between the loads is less than the theoretical differences here, because of degradation in SNR caused by noise added by the receive signal chain.

\begin{table*}[htbp]
    \centering
\resizebox{\textwidth}{!}{%
\begin{tabular}{|c|c|c|c|c|c|c|c|c|}
\hline
\multirow{3}{*}{\textbf{Load}} & \multirow{3}{*}{\textbf{$T_P$ (K)}} & \multirow{3}{*}{\textbf{$T_N$ (K)}} & \multirow{3}{*}{\textbf{$G_{RX}$}} & \multirow{3}{*}{\textbf{B (Hz)}} & \multirow{3}{*}{\textbf{Offset}} & \multirow{3}{*}{\textbf{\begin{tabular}[c]{@{}c@{}}$\widehat{V}^2$ Predicted \\ from Measured $T_P$\end{tabular}}} & \multirow{3}{*}{\textbf{\begin{tabular}[c]{@{}c@{}}$\widehat{V}^2$ Calculated \\ from Extracted $T_N$\end{tabular}}} & \multirow{3}{*}{\textbf{$\widehat{V}^2$ Measured}} \\
 &  &  &  &  &  &  &  &  \\
 &  &  &  &  &  &  &  &  \\ \hline
50$\Omega$ (Room Temperature) & 296 & 296 & \multirow{5}{*}{2.3e11} & \multirow{5}{*}{1e6} & \multirow{5}{*}{0.0212} & 0.0676 & - & 0.0676 \\ \cline{1-3} \cline{7-9} 
50$\Omega$ (Liquid Nitrogen) & 77 & 77 &  &  &  & 0.0332 & - & 0.0333 \\ \cline{1-3} \cline{7-9} 
Open Circuit & 296 & 46 &  &  &  & 0.0217 & - & 0.0274 \\ \cline{1-3} \cline{7-9} 
Short Circuit & 296 & 40 &  &  &  & 0.0221 & - & 0.0283 \\ \cline{1-3} \cline{7-9} 
LNA Input & - & 39.5 &  &  &  & - & 0.0274 & 0.0273 \\ \hline
\end{tabular}%
}
\caption{\textbf{Computing Theoretical Gaussian Distribution.} \normalfont This table shows the parameters used to compute the theoretical $\widehat{V}^2$ for each load. The theoretical $\widehat{V}^2$ values can be compared with the measured $\widehat{V}^2$ in the last column of the table. In the theoretical Gaussian distributions later in the paper, the variable $\widehat{V}^2$ from this table is renamed $\sigma^2$. The $T_P$ values above are measured physical temperatures. The $T_N$ values for the 50$\Omega$ loads are assumed, to produce the linear calibration curve shown in Fig.\ref{fig:linearfit}. The other $T_N$ values in the column are extracted by applying this linear calibration curve to the measured $\widehat{V}^2$ values.} \label{table:v2calc}
\end{table*}

\subsubsection*{(3) Experimental Measurements and Calibration Overview} We measure the noise produced by each load of interest in a cabled benchtop experiment, and use this to compute the corresponding noise temperature along with calibration parameters (i.e., receiver noise offset and gain). In addition to measuring the noise of the 50$\Omega$ load at room temperature, we also evaluate the 50$\Omega$ load when submerged in ice water (273K) and liquid nitrogen (77K). The three different temperatures for the 50$\Omega$ load are considered to be \textit{known} temperatures.  One special load of interest we evaluate in addition to the communication loads is an LNA input, in order to characterize its effective temperature.

\begin{figure}[t!]
    \centering
    \includegraphics[width=1\columnwidth]{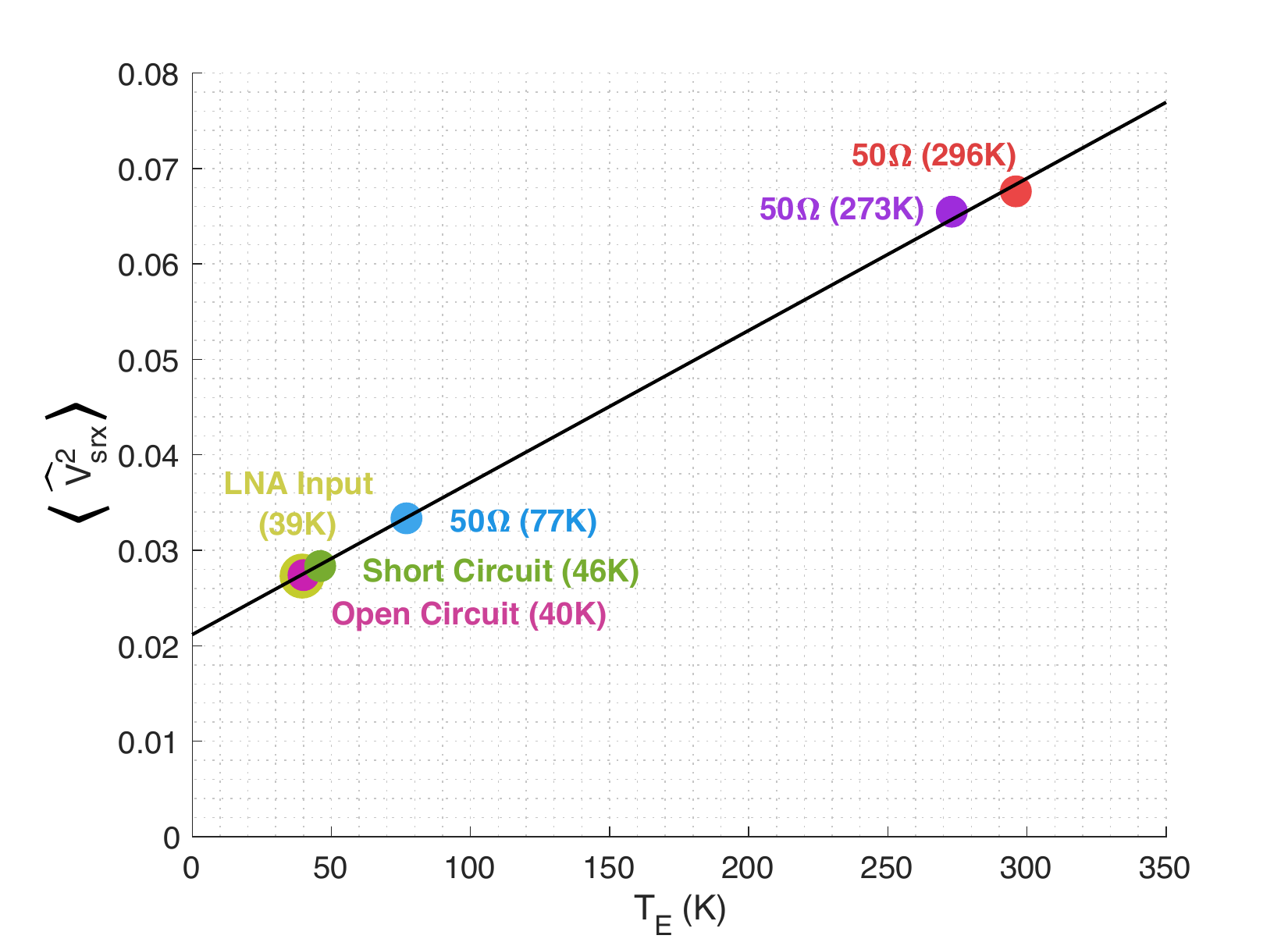}
    \caption{\textbf{Noise Temperature.} The measured mean square noise voltage and noise temperature for each load. The three 50$\Omega$ loads at known temperatures were used to calibrate the noise temperatures for the other loads. Here the linear fit equation is, $\langle\hat{V}_{s_{rx}}^2\rangle = 0.000159\cdot T + 0.0212$}
    \label{fig:linearfit}
\end{figure}

We collected 60 million measurements of the noise generated by each of the loads and found for each  the mean squared voltage $\langle \widehat{V}_{s_{rx}}^2\rangle$ using 

\begin{equation}
    \langle \widehat{V}_{s_{rx}}^2 \rangle = \frac{1}{N}\sum {s_{rx}}^2 
    \label{eq:v2_rx}
\end{equation}
where $s_{rx}$ is the received signal.\footnote{In particular, it is the real part of the complex values provided by the SDR; the imaginary part has the same statistics.} The hat (such as in $\widehat{V}_{s_{rx}}$) indicates that the voltages are in arbitrary SDR units. Voltages without a hat are in physical units (Volts). Using measurements of the 50$\Omega$ load at the three known temperatures, we performed a linear fit of $\langle \widehat{V}_{s_{rx}}^2 \rangle$ versus temperature. Using this linear fit, we extracted the noise temperature of the remaining loads (LNA Input, open, and short) from the observations of $\widehat{V}_{s_{rx}}$. Fig.~\ref{fig:linearfit} shows the results, which indicate that the LNA's noise temperature is cold (39K), much colder than a $50\Omega$ load at room temperature (296K). The open and short circuit loads also appear colder than room temperature. The short circuit terminator has a physical temperature of 296K, but its noise temperature is 46K. The open circuit has a noise temperature of 40K. 

\begin{figure*}[t!]
\centering
\includegraphics[width=1\linewidth]{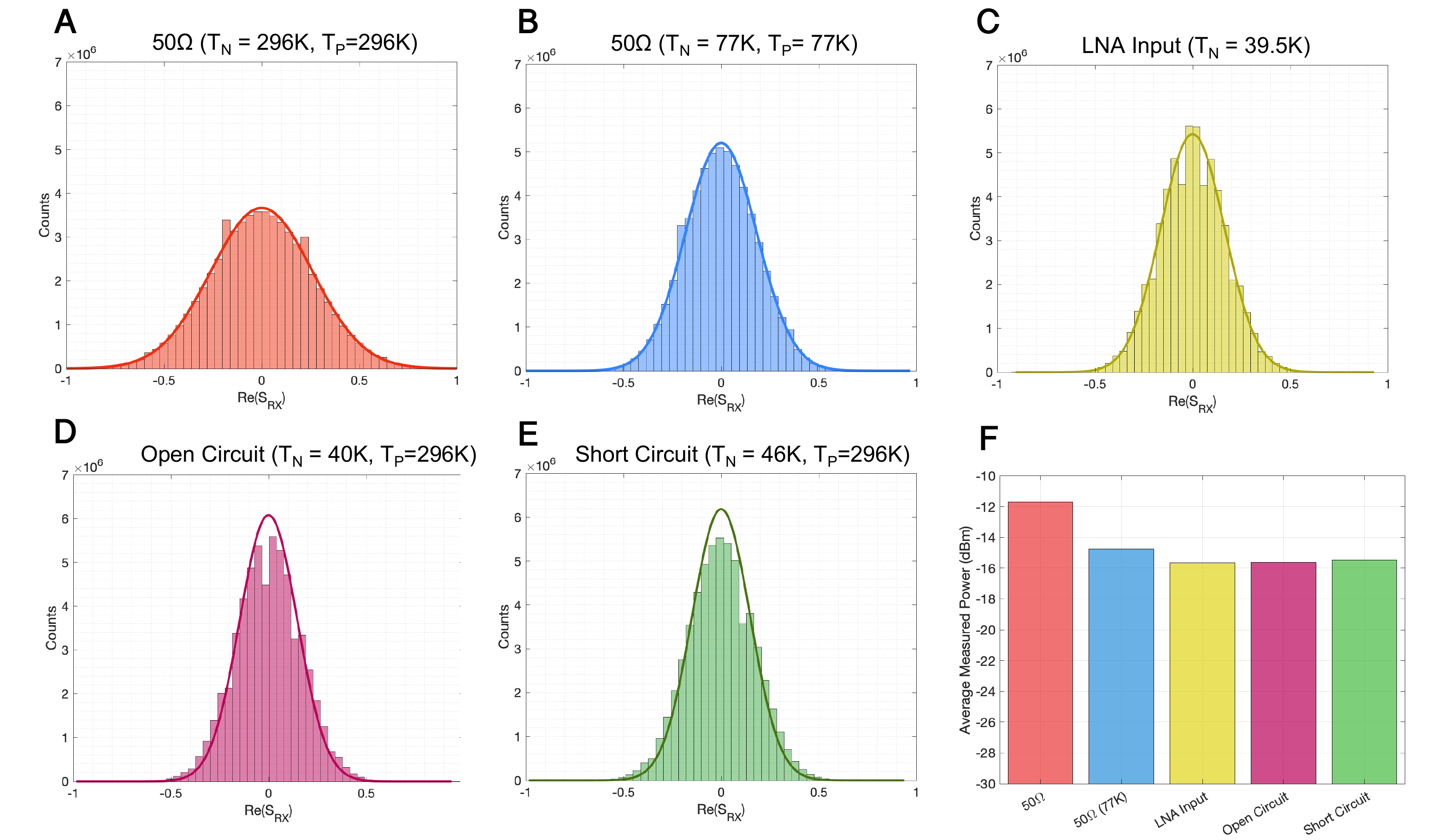}
\caption{\textbf{Theoretical noise distributions compared to data.} For each load, the Noise Temperature $T_N$ and Physical Temperature $T_P$ is indicated at the top of the plot. \textbf{(A):} histogram distribution of the real component of the measured data for the 50$\Omega$ terminator at room temperature (296K). The theoretical Gaussian distribution is overlaid on the data.  \textbf{(B):} 50$\Omega$ terminator in liquid nitrogen (77K)  \textbf{(C):} the LNA input has a noise temperature $T_N$=39K. Its low temperature is important to the operation of the system. \textbf{(D):} the short circuit terminator has a low effective temperature $T_N$=46K because it is poorly matched to the LNA \textbf{(E):} the open circuit terminator has a low temperature of $T_N$=40K, also because of mismatch.  \textbf{(F):} a comparison of the average measured power of each load in dBm after amplification.}
\label{fig:hist}
\end{figure*}

\subsubsection*{(4) Comparing Theoretical and Observed Noise Distributions}
To compare our experimental measurements with expected theoretical values, we also need to perform some calibration. In particular, we need to find the baseline or offset noise produced by the receive chain, as well as the gain value that converts between the SDR's arbitrary units and volts. This calibration of arbitrary SDR units to volts will allow us to compare the theory and experiment for the less obvious loads, namely the open and the short. Using our results from Fig.~\ref{fig:linearfit}, we know that the receive chain introduces noise equivalent to an offset of 0.0212 in arbitrary SDR units. In other words, the offset is the $\langle {\widehat{V}_{s_{rx}}}^2 \rangle$ value at the y-intercept of the linear fit. The gain, which is constant across all measurements, can be computed by determining the ratio of the measured data for any particular load to the expected theoretical value, and then using this same value for all other datasets. The gain is given by,

\begin{equation}
    G_{RX} = \frac{\langle {\widehat{V}_{s_{rx}}}^2 \rangle - offset}{\langle V^2 \rangle}
    \label{eq:gain_rx}
\end{equation}

Now we can compare our theoretical model (based on impedance mismatch of Johnson noise) to measured values. For each load, we compare the histogram of our observed noise measurements with the theoretical distribution.  
To find the distributions, recall that Eq.~\ref{eq:v2_th} predicts $\langle V^2 \rangle$ as a function of Temperature, Bandwidth, Johnson resistor value, and LNA input impedance. Since thermal noise is white, the distribution of the real or imaginary voltages is Gaussian with $\mu = 0$ and $\sigma = \sqrt{\langle V^2 \rangle} = V_{RMS}$. We know from the results in Fig.~\ref{fig:smithchart} that $R_1(short) = 0.254\Omega$ and $R_1(open) = 17k\Omega$ at our frequency of interest. We also must account for the gain and offset of the receive chain. As previously mentioned, the receive chain introduces noise equivalent to an offset of 0.0212 in arbitrary SDR units. We can also compute the receiver gain using Eq.~\ref{eq:gain_rx}. Now, the variance $\sigma^2$ (which we also refer to as $\langle {\widehat{V}_{s_{rx}}}^2 \rangle$) can be related to the theoretical value of $\langle V^2 \rangle$ using 

\begin{equation}
    \sigma^2 = \langle V^2 \rangle \cdot G_{RX} + offset
\end{equation}

The Gaussian distribution is then computed for each case:

\begin{equation}
    f(x) = \frac{1}{\sqrt{2\pi}\sqrt{\sigma^2}}\cdot e^{-\frac{1}{2}\Big(\frac{x-\mu}{\sqrt{\sigma^2}}\Big)^2}
    \label{eq:pdf}
\end{equation}

Fig.~\ref{fig:hist} shows the histogram distribution of the measured data for each load and the the corresponding theoretical distribution. The bandwidth is determined by the bandwidth of the SDR, and in particular its anti-aliasing filters.

\begin{figure*}[th!]
    \centering
    \includegraphics[width=1\textwidth]{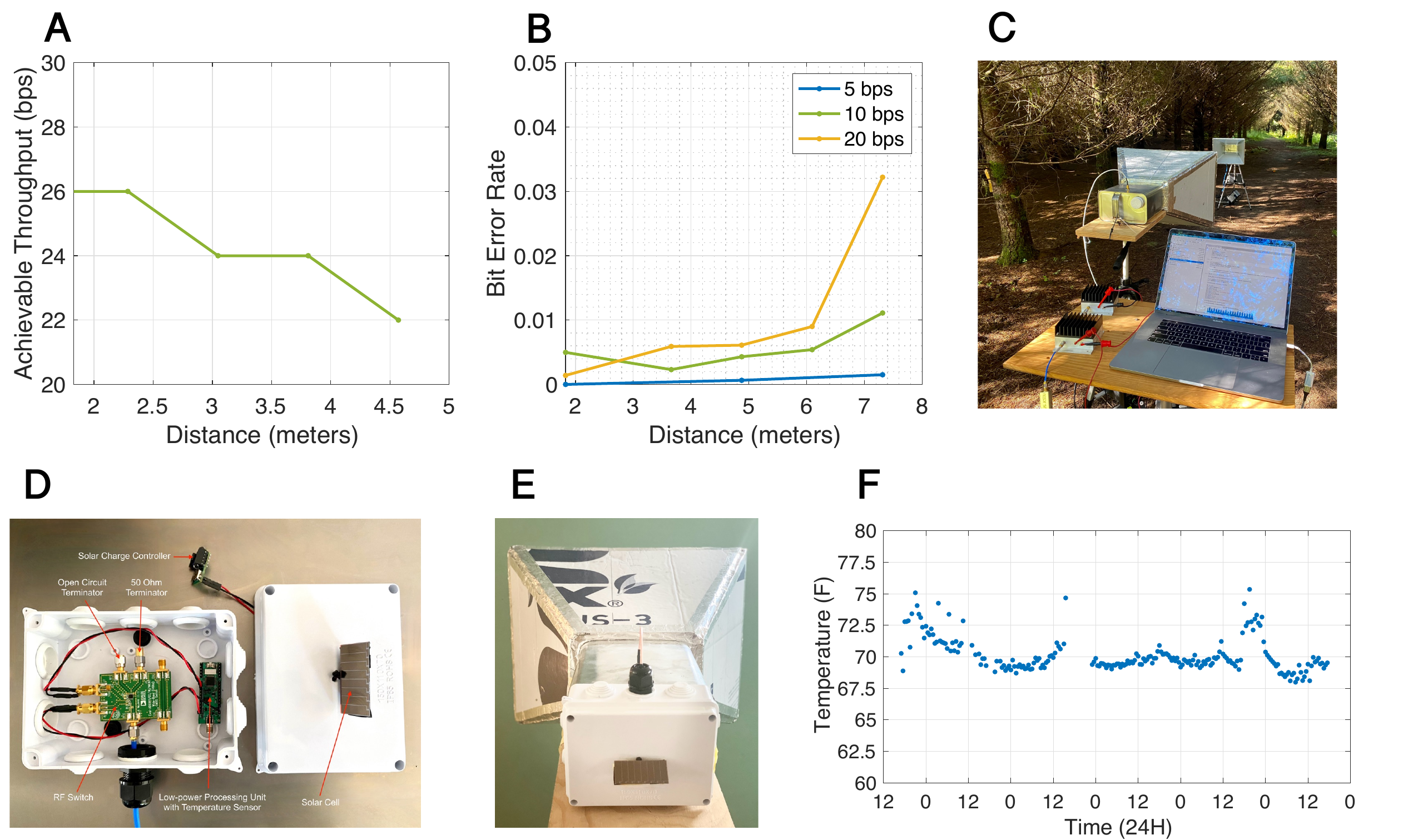}
    \caption{\textbf{Performance Evaluation.} \textbf{(A)} shows the achievable throughput with respect to distance, which was evaluated wirelessly inside of an anechoic chamber. \textbf{(B)} shows the bit-error-rate with respect to distance for three different data rates: 5bps, 10bps, and 20bps. Here, the communication range was evaluated wirelessly in an outdoor setting as shown in \textbf{(C)}. \textbf{(D)} and \textbf{(E)} show a battery-free transmitter prototype that collects ambient temperature data and transmits the information by means of modulated Johnson noise. The temperature plot in \textbf{(F)} shows the data that was collected wirelessly across 5 consecutive days in a residential building.}
    \label{fig:performance}
\end{figure*}

\section*{Performance} 
The performance of the wireless communication system is evaluated in terms of achievable throughput and communication range by conducting wireless experiments using the previously described hardware. Additionally, we demonstrate the potential for battery-free sensing by implementing a battery-free transmitter that transmits data by modulating Johnson noise.

\subsection*{Achievable Throughput} To evaluate achievable throughput, the measurements were performed inside of an anechoic chamber. This allows us to test the system in an interference free environment and to achieve best possible performance given our hardware implementation and the modulation and demodulation scheme. Measurements were taken every 1.5 meters and up to 4.5 meters inside the anechoic chamber, and at each distance data packets were transmitted wirelessly. The data packet structure includes a 7-bit Barker code as the preamble, followed by a 13-bit data payload. At each distance, the maximum achievable throughput was determined by measuring which data rate would allow us to achieve a bit-error-rate (BER) less than or equal to 1\%. Fig.~\ref{fig:performance}\textcolor{blue}{A} shows that the maximum achievable throughput is 26bps at 1.5 meters and goes down to 22bps at 4.5 meters. Moreover, the number of samples per bit was fixed to 5 while the subcarrier frequency was varied, which in turn varies the data rate. For example, at 2.3 meters the achievable throughput was 26bps and data was modulated by using a 130Hz subcarrier frequency. At 4.6 meters the achievable throughput was 22bps and data was modulated by using a 110Hz subcarrier frequency.

\subsection*{Communication Range} The wireless communication range is evaluated in an outdoor environment, as shown in Fig.~\ref{fig:performance}\textcolor{blue}{C}. Here, measurements were taken starting from 1.8 meters and up to 7.3 meters for three different data rates. We evaluate the communication range using data rates of 5, 10, and 20bps. Fig.~\ref{fig:performance}\textcolor{blue}{B} shows the range performance results. As expected, a higher data rate results in a short communication range. For example, with a data rate of 20bps the maximum communication range is approximately 6.1 meters. Whereas, using a data rate of 5bps the communication range increases to approximately 7.3 meters while having a BER = 0.15\%. 

\subsection*{Battery-free Transmitter}
We demonstrate the potential for battery-free sensing using  modulated Johnson noise, by developing a battery-free transmitter, as shown in Fig.~\ref{fig:performance}\textcolor{blue}{D} and \textcolor{blue}{E}. The transmitter includes an ADG901 RF switch and a custom PCB that integrates a low-power MCU, temperature sensor, and solar energy harvester~\cite{multiscatter}. The custom PCB is used to collect temperature data, packetize the data, and control the RF switch to modulate data. The transmitter is powered using a small 2.54cm. x 6.35cm. solar cell that charges a supercapacitor. We deploy the battery-free transmitter inside a residential building to monitor ambient temperature data and transmit the information by means of modulated Johnson noise. The data was received using the receiver hardware previously described. In Fig.~\ref{fig:performance}\textcolor{blue}{F}, we show the temperature collected across five days. Note that no error correcting codes were implemented which resulted in some anomalies in the ambient temperature data collection. 

\section*{Conclusion}
This paper opens up a new direction for low-power wireless communication by demonstrating for the first time that information bits can be wirelessly transmitted by modulating Johnson noise. By selectively connecting and disconnecting an impedance matched resistor to an antenna, data can be wirelessly transmitted with no active oscillator on either the transmit or the receive side, and no pre-existing RF carrier, as in backscatter and ambient backscatter communication. We have demonstrated that our system can achieve data rates up to 26bps and communication range as far as 7.3 meters. This method has many attractive features compared to previous passive communication schemes such as ambient backscatter or RFID, because it is not reliant on ambient or generated RF sources. One of the challenges faced by backscatter communication is the increased system complexity: conventional half-duplex radio transmission requires just two entities, a transmitter and a receiver. Backscatter requires three entities because it needs a carrier generator in addition to the data transmitter and receiver. (In a monostatic backscatter reader, the carrier generator is packaged with the receiver, but the system architecture is still more complex.) In addition to the deployment benefits of the simplified system architecture, the absence of a carrier means that the reader device does not need to contend with self-jamming caused by the carrier, which is a key driver of RFID reader complexity and cost. While the performance of the prototype system is modest, we hope that the techniques presented will lead to a new avenue of research and help realize the vision of ubiquitous computing.

\acknow{We thank A. Saffari and D. Nissanka for assisting in the prototype development, S. Garman for performing antenna characterization, and M. Reynolds for suggesting the two temperatures 50$\Omega$ experiment.
\newline
A provisional patent has been filed on the subject matter with the USPTO.}

\showacknow{} 

\bibliography{main}

\section*{Supporting Information Appendix (SI)}
 
\subsection*{Sytem Validation}
We provided additional details of the system validation in the following sections.

\subsubsection*{Experimental Setup}
\begin{figure*}[h!]
\centering
\includegraphics[width=1\textwidth]{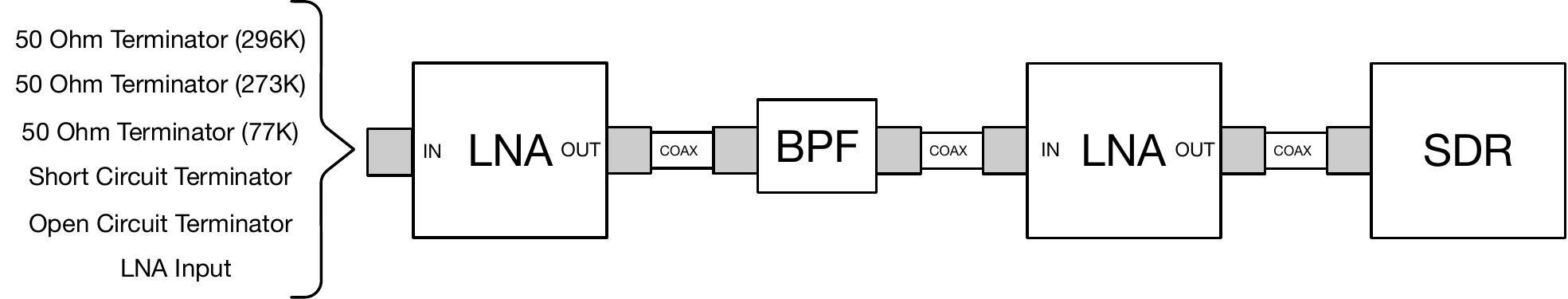}
\caption{\textbf{Noise Measurements Experiment.} A block diagram of the cabled experimental setup to evaluate the noise temperature and noise power of various loads.}
\label{fig:setup}
\end{figure*}

Fig.~\ref{fig:setup} shows the experimental setup used to collect data of several different loads connected to th receiver: 50$\Omega$ at three different temperatures, short circuit, open circuit, and the input of an LNA. To clarify the meaning of "LNA input" as a load: we take an additional powered up LNA and connect its input to the input of our standard receive signal chain (which consists of two more LNAs, bandpass filter, etc.). The LNA that is used as a load is identical to the LNAs used in the receive chain. The purpose of measuring the LNA input as a load is to measure its effective noise temperature, which will determine the baseline noise that the received communication signals will be added to, and will also determine the properties of the radiation field that the receive LNA emits back toward the transmitter. 

\begin{figure*}[h!]
\centering
\includegraphics[width=0.7\textwidth]{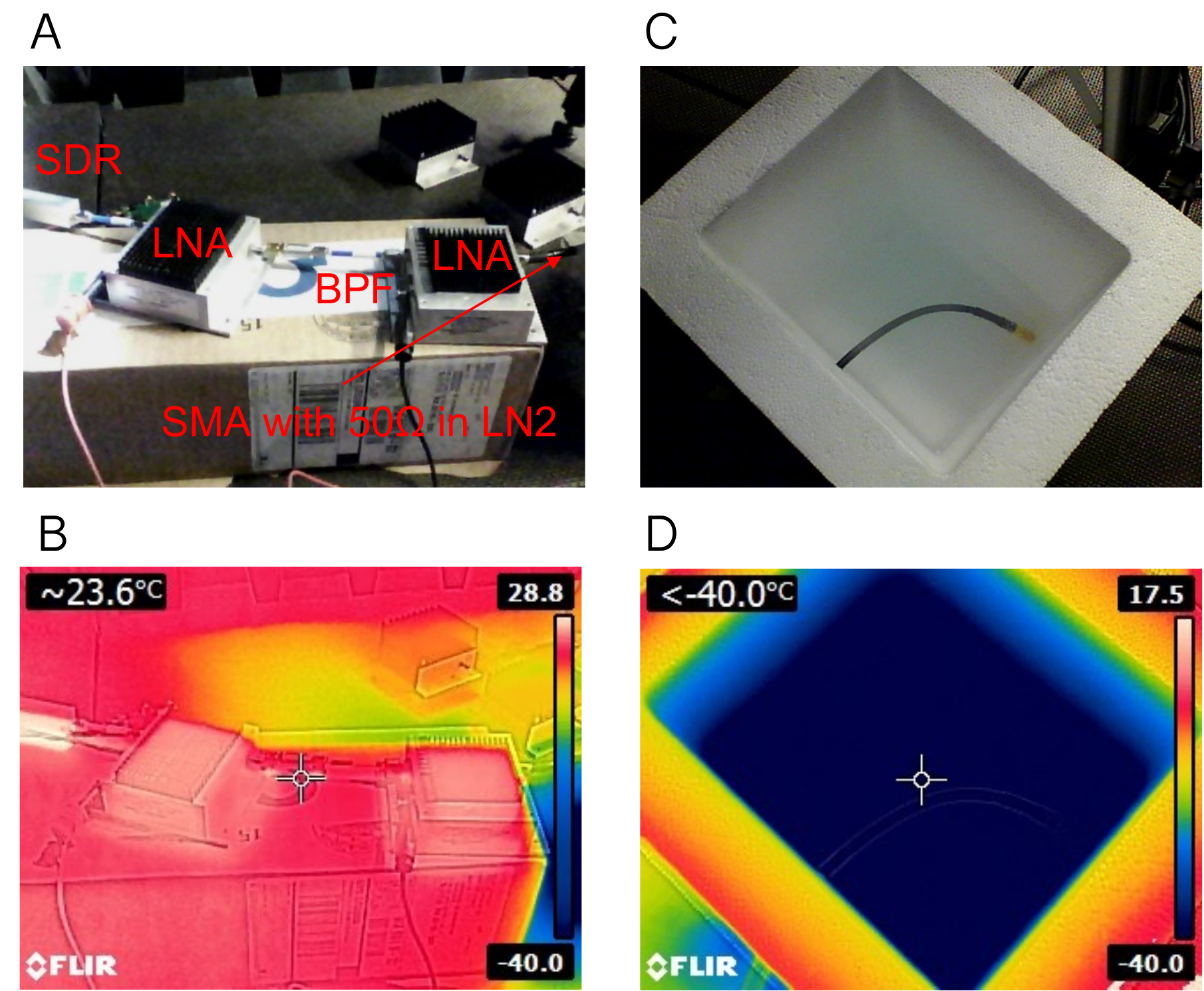}
\caption{\textbf{Experimental Setup.} \textbf{(A)} The cabled experimental setup inside of an anechoic chamber and (B) shows the corresponding thermal image. The 50$\Omega$ load submerged in liquid nitrogen is shown in \textbf{(C)} and the corresponding thermal image is shown in \textbf{(D)}. Note that the measurement range of the thermal imaging camera does not reach liquid nitrogen temperatures.}
\label{fig:thermal}
\end{figure*}

To lower the temperature of the 50$\Omega$ load to 273K and 77K, we submerged the load in ice water and liquid nitrogen, respectively. The temperature values for room temperature and ice water were measured with a thermal camera. The liquid nitrogen temperature was taken to be the standard value for liquid nitrogen (77K) because its temperature was below the range of the thermal camera. In Fig.~\ref{fig:thermal} we show the thermal images of the cabled experimental setup.

\subsection*{Performance Evaluation}
To evaluate the performance of our prototype implementation we used off-the-shelf hardware. The transmitter includes an ADG901 RF Switch, a 50$\Omega$ terminator, an open circuit terminator, and a Raspberry Pi 3B was used to control the RF switch. The receiver includes two ZHL-1217HLN+ LNAs, a VBF-1445+ bandpass filter, an RTL-SDR dongle, and a laptop PC ~\cite{lna, bpf, SDR}. Both the transmitter and receiver use horn antennas that were constructed using readily available materials and were based off of the designs provided by the West Virginia University Radio Astronomy Laboratory~\cite{horn}. 

\subsubsection*{SDR Hardware Configuration}
We note that the SDR must have a \textit{fixed} gain value configured. In particular, when working with RTL-SDR devices, the gain is not set to a fixed value by default. Not doing so can results in inconsistent data recordings. Additionally, it is common for certain RTL-SDR devices to produce a constant DC 'spike'. In Fig. \textcolor{blue}{1A} of the main text, a Hampel filter was used to remove the DC spike, specifically with number of neighbors, k = 100. Lastly, to interface with the SDR, the Python RTL-SDR library was used~\cite{pyrtlsdr}. 

\subsection*{Power Consumption Comparison}
We compare the power consumption of our transmitter to other passive and low-power wireless communication techniques as shown in Table~\ref{table:pwr}. The transmitter power consumption is identical to that of ambient backscatter solutions and is less power consuming than traditional RFID (WISP platform). We also look into low-power BLE solutions (Atmosic ATM33) which is more power consuming than Modulated Johnson Noise and the backscatter solutions. Of course, the achievable throughput must also be considered, and other solutions would outperform modulated Johnson noise in this regard. For example, ambient backscatter has shown to achieve data rates of up to 10kbps, the WISP can achieve data rates of up to 256kbps, and the Atmosic ATM33 can achieve data rates ranging from 125kbps - 2Mbps~\cite{ambientbackscatter, wisp, atmosic}. While Modulated Johnson Noise is limited to low data rates, a key benefit is that it is not dependent on ambient or generated RF signals. 

\newpage

\begin{table*}[h!]
\centering
\begin{tabular}{|c|c|}
\hline
\textbf{Solution}       & \textbf{Power Consumption (TX)} \\ \hline
Modulated Johnson Noise & 0.25 uW                         \\ \hline
Ambient Backscatter~\cite{ambientbackscatter}     & 0.25 uW \\ \hline
RFID~\cite{wisp}                    & 2.32 uW                         \\ \hline
Low-Power BLE~\cite{atmosic}           &     6.3 mW                     \\ \hline
\end{tabular}
\caption{\textbf{Power Consumption Comparison.} \normalfont A comparison of the power consumption of Modulated Johnson Noise to existing low-power wireless communication solutions. }
\label{table:pwr}
\end{table*}

\end{document}